\documentclass[11pt]{article} 

\usepackage{floatrow}
\newfloatcommand{capbtabbox}{table}[][\FBwidth]

\usepackage{fullpage} 
\usepackage{graphicx}
\usepackage{subcaption}
\usepackage{upref}
\usepackage{enumerate}
\usepackage{latexsym}
\usepackage{color,graphics}
\usepackage{relsize}
\usepackage{float}
\restylefloat{table}
\usepackage{algorithm}
\usepackage[redeflists]{IEEEtrantools}
\usepackage{centernot}
\usepackage{mathtools}
\usepackage{stmaryrd}
\usepackage{amsmath,amsthm,amssymb,algorithmic,epsfig,graphicx, tikz}
\usepackage{amsfonts}
\usepackage{varioref}
\usepackage[ansinew]{inputenc}
\usepackage{qcircuit}
\usepackage{amsmath}
\usepackage{amsfonts}
\usepackage{tikz} 
\usepackage{multirow}
\usepackage{bbm}
\usepackage{amssymb}
\usepackage{varioref}
\usepackage[ansinew]{inputenc}
\usepackage{authblk}
\usepackage{multicol}
\usepackage{textcomp}
\usepackage{booktabs}
\usepackage{multirow}

\newcommand{\norm}[1]{\left\lVert#1\right\rVert}
\newtheorem{theorem}{Theorem}[section]
\newtheorem{lemma}[theorem]{Lemma}
\newtheorem{claim}[theorem]{Claim}

\newtheorem{assumption*}{Assumption}
\newtheorem{definition}{Definition}
\newcommand{\R}{\mathbb{R}}

\def\bra#1{\mathinner{\langle{#1}|}}
\def\ket#1{\mathinner{|{#1}\rangle}}

\renewcommand{\part}[2]{\frac{\partial #1}{\partial #2}}

\makeatletter
\newcommand{\xMapsto}[2][]{\ext@arrow 0599{\Mapstofill@}{#1}{#2}}
\def\Mapstofill@{\arrowfill@{\Mapstochar\Relbar}\Relbar\Rightarrow}
\makeatother

\usepackage{hyperref}

\newcommand{\thmref}[1]{\hyperref[#1]{{Theorem~\ref*{#1}}}}
\newcommand{\lemref}[1]{\hyperref[#1]{{Lemma~\ref*{#1}}}}
\newcommand{\remref}[1]{\hyperref[#1]{{Remark~\ref*{#1}}}}
\newcommand{\corref}[1]{\hyperref[#1]{{Corollary~\ref*{#1}}}}
\newcommand{\eqnref}[1]{\hyperref[#1]{{Equation~(\ref*{#1})}}}
\newcommand{\claimref}[1]{\hyperref[#1]{{Claim~\ref*{#1}}}}
\newcommand{\remarkref}[1]{\hyperref[#1]{{Remark~\ref*{#1}}}}
\newcommand{\propref}[1]{\hyperref[#1]{{Proposition~\ref*{#1}}}}
\newcommand{\factref}[1]{\hyperref[#1]{{Fact~\ref*{#1}}}}
\newcommand{\defref}[1]{\hyperref[#1]{{Definition~\ref*{#1}}}}
\newcommand{\exampleref}[1]{\hyperref[#1]{{Example~\ref*{#1}}}}
\newcommand{\hypref}[1]{\hyperref[#1]{{Hypothesis~\ref*{#1}}}}
\newcommand{\secref}[1]{\hyperref[#1]{{Section~\ref*{#1}}}}
\newcommand{\chapref}[1]{\hyperref[#1]{{Chapter~\ref*{#1}}}}
\newcommand{\apref}[1]{\hyperref[#1]{{Appendix~\ref*{#1}}}}

\newcommand\blfootnote[1]{
  \begingroup
  \renewcommand\thefootnote{}\footnote{#1}
  \addtocounter{footnote}{-1}
  \endgroup
}

\AtBeginDocument{}

\begin{document}
\bstctlcite{IEEEexample:BSTcontrol}

\title{Quantum Spectral Clustering}

\author[]{Iordanis Kerenidis}
\author[]{Jonas Landman}
\affil[]{Universit\'e de Paris, IRIF, CNRS, Paris, France}

\maketitle
\begin{abstract}
Spectral clustering is a powerful unsupervised machine learning algorithm for clustering data with non convex or nested structures \cite{ng2002spectral}. With roots in graph theory, it uses the spectral properties of the Laplacian matrix to project the data in a low-dimensional space where clustering is more efficient. Despite its success in clustering tasks, spectral clustering suffers in practice from a fast-growing running time of $O(n^3)$, where $n$ is the number of points in the dataset. In this work we propose an end-to-end quantum algorithm performing spectral clustering, extending a number of works in quantum machine learning. 
The quantum algorithm is composed of two parts: the first is the efficient creation of the quantum state corresponding to the projected Laplacian matrix, and the second consists of applying the existing quantum analogue of the $k$-means algorithm \cite{qmeans}. 
Both steps depend polynomially on the number of clusters, as well as precision and data parameters arising from quantum procedures, and polylogarithmically on the dimension of the input vectors. Our numerical simulations show an asymptotic linear growth with $n$ when all terms are taken into account, significantly better than the classical cubic growth.
This work opens the path to other graph-based quantum machine learning algorithms, as it provides routines for efficient computation and quantum access to the Incidence, Adjacency, and projected Laplacian matrices of a graph.

\end{abstract}

\blfootnote{Email: landman@irif.fr}

\section{Introduction}

We have recently witnessed extraordinary practical advances in quantum computing \cite{quantumsupremacy} and the emergence of several quantum algorithms for machine learning with promising potential speedups compared to their classical analogues \cite{lloyd2014quantum, wiebe2014quantum, aimeur2013quantum, WKS14, KP16, childs2017quantum, QCNN}. These algorithms show the potential of full-scale quantum computers and can also provide new insight on how to find quantum applications of quantum computing in the near future. 

In this work, we develop an end-to-end quantum algorithm for Spectral Clustering. Introduced in \cite{ng2002spectral}, this unsupervised learning algorithm shows great accuracy in identifying complex and interlacing clusters, and allows a high level of explainability. However, it suffers from a fast-growing runtime, namely cubic in the number $n$ of vectors in the dataset, that inhibits its use in practice. The goal of the Spectral Clustering algorithm is to perform the clustering task in a low-dimensional space derived from the data. More precisely, one starts with the set of input vectors and constructs a similarity graph, where the edge between two nodes is built from the distance between the two associated vectors. From this graph, one can define the Incidence matrix and the Laplacian matrix. By extracting the eigenvectors of the Laplacian matrix and keeping the subspace spanned by the lowest ones, a lower-dimensional space is defined. The initial data can then be projected onto this new space, where one can expect the data to be better organized, as clusters. Therefore, to obtain the clusters, the $k$-means algorithm is applied on the projected vectors.

\paragraph{Result}

We provide a quantum algorithm for spectral clustering following a similar methodology, while having to carefully extend or alter the specifics of each step of the algorithm. We first adapt and extend recent and efficient quantum subroutines for linear algebra and distance estimation. These methods are used to create the similarity graph, as well as the Incidence and Laplacian matrices, extract their eigenvectors, and project the data points onto the right subspace to finally apply a quantum analogue of $k$-means. While the steps of the algorithm follow the classical ones, we had to adjust most of the definitions, for example the definitions of the Incidence and Laplacian matrices, in order to make the quantum algorithm efficient. We will detail all the needed changes in the following sections.

In high level, the running time of the quantum spectral clustering algorithm reflects the two stages of spectral clustering and is given by
\begin{equation}
O( T_{\widetilde{\mathcal{L}}^{(k)}} T_{qmeans})
\end{equation} 
The first term $T_{\widetilde{\mathcal{L}}^{(k)}}$ is the time to create a quantum state corresponding to the normalized Laplacian matrix of the graph projected on its lowest eigenvectors. The resulting quantum state is the input state to the quantum clustering algorithm whose overall running time contains another multiplicative term that we denote by  $T_{qmeans}$.

For the first part, we will propose an algorithm in time 
\begin{equation}
T_{\widetilde{\mathcal{L}}^{(k)}}= \widetilde{O}\left( T_S 
    \frac{\eta(S)}{\epsilon_{dist}\epsilon_B} 
    \frac{\mu(\mathcal{B})\kappa(\widetilde{\mathcal{L}}^{(k)})}{\epsilon_{\lambda}} \right).
\end{equation}
Here, the term $T_S$ is the time to load the input vectors in a quantum state, which becomes efficient if we assume quantum memory (QRAM) access (see discussion after Definition \ref{definitionquantumaccess}). The terms $\epsilon_{dist},\epsilon_B$, and $ \epsilon_{\lambda}$ correspond to error or precision parameters that appear in several quantum routines. The matrices $S, \mathcal{B},$ and $\widetilde{\mathcal{L}}^{(k)}$ are respectively the input data matrix, the normalized incidence matrix, and the projection of the normalized Laplacian matrix. For these matrices, the condition number $\kappa(\cdot)$ is the ratio between the largest and the lowest singular values, and the terms $\mu(\cdot)$ and $\eta(\cdot)$ are two specific norm parameters defined respectively in Definitions \ref{definitionmu} and \ref{etadefinition} (see Section \ref{quantumpreliminaries} for more details).

We will show that the term $\mu(B)$ in the above expression is in fact upper bounded by $O(n)$, in the worst case scenario. In our basic numerical experiments (see Section \ref{NumericalSimulations}), we indeed observed a quantum running time scaling linearly with $n$, when all terms are taken into account. This implies a significant polynomial speedup over the classical algorithm.
    
The second stage, the quantum clustering, adds to the running time the multiplicative term $T_{qmeans}$, a rather complex expression detailed in \cite{qmeans} (see also Theorem \ref{qmeansthm}). In the case of \emph{well-clusterability}, namely when the vectors can effectively be organized in clusters that can be detected efficiently (See Definition \ref{wellclusterabilitydef} for details), this runtime can be rewritten as follows, where $k$ is the number of clusters and $\delta$ is a precision parameter : 
\begin{equation}
T_{qmeans} = \widetilde{O}\left(\frac{k^3\eta({\widetilde{\mathcal{L}}^{(k)}})^{2.5}}{\delta^3}\right)
\end{equation}

We expect our quantum spectral clustering algorithm to be accurate and efficient when the classical spectral clustering algorithm also works well. In fact, the classical algorithm works well in the case when, once projected onto the reduced spectral space, the vectors follow the \emph{well-clusterability} assumption, allowing for efficient clustering. In particular, in this case, the term $\kappa(\widetilde{\mathcal{L}}^{(k)})$ is close to $k$, the number of clusters. Note however that our algorithm could work without this \emph{well-clusterability} assumption, but the theoretical runtime would be bounded differently. 

In conclusion, our algorithm provides a considerable theoretical speedup that could allow for new applications of Spectral Clustering on larger, high-dimensional datasets. The quantum subroutines developed in this paper could be useful independently, and we hope for substantial improvements in several other graph-based machine learning algorithms.

\paragraph{Related Work}

There is extensive work in quantum computing involving graph problems such as min-cut, max-flow, or the traveling salesman problem \cite{moylett2017quantum,maxcut_cui16}, but only a few are about graph-based machine learning \cite{Otterbach17, drineas2004clustering}. Spectral clustering has been studied in \cite{daskin2017quantum} but no proven speedups were given.
More recently, \cite{apers2019quantum} introduced quantum algorithms using the graph Laplacian for optimization and machine learning applications, including spectral clustering. In that paper, the starting point was the assumption of having superposition access to the classically stored weights of the similarity graph, corresponding to the Laplacian directly, from which the authors performed tasks like sparsification of the graph faster than classical algorithms. In our work, we propose an efficient quantum algorithm to construct the projected Laplacian matrix itself from access to the classical input, before proceeding to the clustering algorithm. Note that one could eventually combine the methods from \cite{apers2019quantum} and our procedure, or any other procedure that provides access to the projected Laplacian matrix, and thus construct different Quantum Spectral Clustering algorithms.\\

The remainder of the paper is organized as follows: In Section \ref{classicalalgo} we formulate and analyze the classical spectral clustering procedures, introducing all necessary notation for the classical and quantum case. In Section \ref{quantumpreliminaries} we present the necessary background in quantum information as well as theorems and definitions in quantum machine learning that will be used in our work. In Section \ref{quantumspectralclusteringsection} we present our algorithm in detail and perform an error and runtime analysis. Finally, numerical simulations are reported in Section \ref{NumericalSimulations} to validate the performance of our algorithm in practice.

\section{Classical Spectral Clustering}\label{classicalalgo}

\subsection{Notation and Definitions}\label{notationsanddefinition}

We use the following notation: $\norm{\cdot}$ is the $\ell_2$ norm of a vector, $\norm{\cdot}_F$ is the Frobenius norm for a matrix, $\widetilde{O}()$ symbolizes algorithm complexity or running time where parameters with polylogarithmic dependence are hidden. 
The condition number $\kappa(M)$ of a matrix $M$ is defined as the ratio between its largest and its smallest non zero eigenvalues (or singular values). $M^T$ is the transpose of $M$.\\

\begin{figure}[h!]
\centering
   \includegraphics[width=\textwidth]{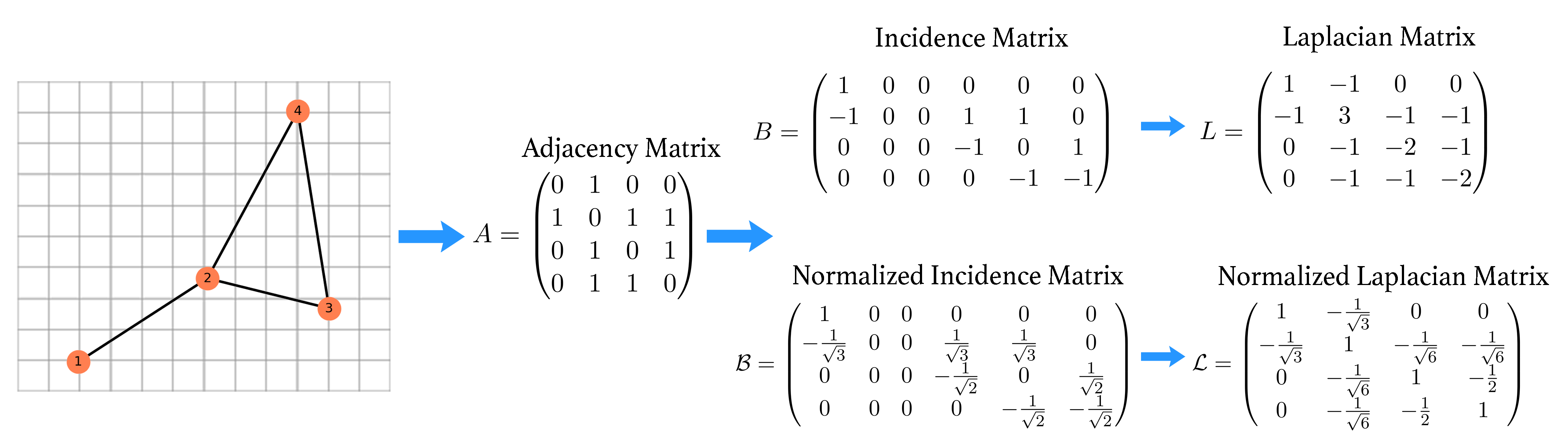}
\caption{a 4 nodes graph}
\label{graph}
\end{figure}

%

Let $S \in \R^{n\times d}$ be the input of our clustering task (see Table \ref{variable_summary}). $S$ is the data matrix composed of $n$ vectors $s_i \in \R^{d}$, for $i \in[n]$. The spectral clustering method uses a graph derived from the data $S$, where similar points are connected. We define the distance between two points by $d_{ij}=\norm{s_i-s_j}$.\\

We consider the undirected graph for which each of the $n$ nodes corresponds to a data point.
The value of the edge connecting two nodes $i$ and $j$ is 1 if the two nodes are connected and 0 otherwise. More generally we will note $a_{ij} \in \{0,1\}$ the value of this edge. By convention we have $a_{ii}=0$. 
We define the Adjacency matrix $A \in\R^{n\times n}$ as the symmetric matrix with elements $a_{ij}$. \\

We will use the following construction rule for the graph: the value of an edge between two points $s_i$ and $s_j$ is equal to 1 if their distance satisfies $d_{ij} \leq d_{min}$ and 0 otherwise, for a given threshold $d_{min} >0$. This choice has been made for simplicity and to take into account constraints from quantum circuits that will be detailed later. \\

The Incidence matrix $B$ is another way of representing the graph. Each row of $B$ represents a node whereas a column represents a possible edge. An element of $B$ indicates whether an edge is incident to a node. $B$ is not symmetric and has size $n \times \frac{n(n-1)}{2}$. 
We index the elements of $B$ by three numbers $B_{i,(p,q)}$ where $i$ is the node and $(p,q)$ represents the edge connecting the nodes $p$ and $q$ ordered so that $p<q$. Even though the graph is undirected, the values of $B$ must follow an oriented convention. Therefore the rule for constructing $B$ is the following:
\begin{equation}\label{Bdefinition}
    B_{i,(p,q)} = \begin{cases}
    a_{pq} \quad \text{if } i=p \\
    -a_{pq} \text{ if } i=q \\
    0 \qquad \text{ if } i\notin \{p,q\}  
    \end{cases}
\end{equation}


We introduce the \emph{normalized} incidence matrix $\mathcal{B}$, with elements defined by $\mathcal{B}_{i,(p,q)}=\frac{B_{i,(p,q)}}{\norm{B_i}}$, where $B_i$ is the $i^{th}$ row of $B$. Therefore each row $\mathcal{B}_i$ has unit norm.\\

\begin{table}[t!]
\centering
\begin{tabular}{|l|c|c|l|}
\hline
\multicolumn{1}{|c|}{Variable}     & Notation                        & Dimension                  & \multicolumn{1}{c|}{Remark}                                                                                                                                                        \\ \hline
Input Matrix                   & $S$                             & $n\times d$                & Each input vector is a row $s_i \in \R^d$                                                                                                                                                                     \\ \hline
Adjacency Matrix               & $A$                             & $n\times n$                & \begin{tabular}[c]{@{}l@{}}Nodes connectivity. Elements $a_{ij}$ = \\ $1$ if $d(s_i,s_j)\leq d_{min}$ and 0 otherwise.\end{tabular}                                                                                                                                   \\ \hline
Incidence Matrix               & $B$                             & $n\times \frac{n(n-1)}{2}$ & \begin{tabular}[c]{@{}l@{}}Elements $B_{i,(p,q)}$ =  \\ $\pm a_{pq}$ if edge $(p,q)$ is incident to node $i$\end{tabular}                                                         \\ \hline
Normalized Incidence Matrix    & $\mathcal{B}$                   & $n\times \frac{n(n-1)}{2}$ & \begin{tabular}[c]{@{}l@{}}Each row has been divided by its norm \\ Eigenvalues $\lambda_1^{\mathcal{B}}\leq\cdots\leq \lambda_n^{\mathcal{B}}$\end{tabular}                      \\ \hline
Normalized Laplacian           & $\mathcal{L}$                   & $n\times n$                & \begin{tabular}[c]{@{}l@{}}$\mathcal{L} = \mathcal{B}\mathcal{B}^T$\\ Eigenvalues $\lambda_j = (\lambda_j^{\mathcal{B}})^2$\end{tabular} \\ \hline
Projected Normalized Laplacian & $\widetilde{\mathcal{L}}^{(k)}$ & $n\times k$                & $\mathcal{L}$ projected on its $k$ lowest eigenvectors                                                                                                                                \\ \hline
\end{tabular}
\caption{Summary of variables for Classical and Quantum Spectral Clustering.}
\label{variable_summary}
\end{table}

The Laplacian matrix is defined by $L=BB^T$. We introduce the normalized Laplacian matrix as $\mathcal{L} = \mathcal{B}\mathcal{B}^T$. It inherits the properties of the Laplacian matrix and will be used for classification. 
Note that the usual definition $\mathcal{L} = D^{-\frac{1}{2}}LD^{-\frac{1}{2}}$, with $D$ the Degree matrix, coincides if the edges are either 0 or 1. \\

$\mathcal{L}$ is a symmetric and positive-semidefinite matrix in $\R^{n\times n}$. The $n$ eigenvalues of $\mathcal{L}$ are real and positive. We note them $\{\lambda_1, \cdots, \lambda_n\}$, and their corresponding eigenvectors are $\{u_1, \cdots, u_n\}$. The eigenvalues are ordered such that $\lambda_1 \leq \lambda_2 \leq \cdots \leq \lambda_n $. 
For a given integer $k \in [n]$, we will note $\widetilde{\mathcal{L}}^{(k)}$ 
the projection of $\mathcal{L}$ on its $k$ lowest eigenvalues.\\

Since $\mathcal{L}=\mathcal{B}\mathcal{B}^T$, the $n$ singular values $\lambda^\mathcal{B}_j$ of $\mathcal{B}$ are such that $\lambda_j = (\lambda^\mathcal{B}_j)^2$. Indeed, using the singular value decomposition (SVD), there exist two orthonormal matrices $U$ and $V$ and the diagonal matrix $\Sigma \in \R^{n \times n}$ with elements $(\lambda^\mathcal{B}_1,\cdots,\lambda^\mathcal{B}_n)$, such that $\mathcal{B}= U\Sigma V^T$. Therefore $\mathcal{L}=U\Sigma^2 U^T$, and the eigenvectors $u_j$ of $\mathcal{L}$ are the left singular vectors of $\mathcal{B}$.\\

Later, to ensure a better running time, we will slightly modify the incidence matrix $B$ by replacing the ``0'' elements by a small parameter $\epsilon_B > 0$. We will see in the experiments that it does not affect the accuracy of the clustering. 

\subsection{Partitioning the Graph into $k$ Clusters}\label{classicalkpartition}
Once the graph's normalized incidence matrix $\mathcal{B}$ is computed, we can calculate its normalized Laplacian and find its eigenvalues and eigenvectors.

Let $\widetilde{\mathcal{L}}^{(k)} \in \R^{n\times k}$ be the \emph{projected} normalized Laplacian matrix on its $k$ lowest eigenvectors; i.e., the $j^{th}$ column of $\widetilde{\mathcal{L}}^{(k)}$ is $u_j$, the $j^{th}$ eigenvector of $\mathcal{L}$, for $j=1,\cdots, k$. 
The method developed by \cite{ng2002spectral} consists of applying the clustering algorithm $k$-means with input the $n$ rows $\widetilde{\mathcal{L}}^{(k)}_i$ of the projected normalized Laplacian $\widetilde{\mathcal{L}}^{(k)}$. 
Each row $\widetilde{\mathcal{L}}^{(k)}_i$ is a vector of dimension $k$, corresponding to an input vector $s_i$ in the initial input space.

With this procedure, the $k$-means clustering takes place in a low-dimensional and appropriate space, ensuring an efficient clustering. 
The output of the algorithm could be the label of each point (for example the label corresponding to the nearest centroid) or the $k$ centroids in the spectral space.

\subsection{Classical Running Time}

The classical algorithm can be decomposed in several steps: the distance calculation between points runs in time $O(dn^2)$, and the creation of the Laplacian matrix in $O(nm)$ where $m$ is the number of edges in the graph, which is $O(n^2)$ in the worst case. Then, the extraction of eigenvalue and eigenvectors of the Laplacian matrix is done in $O(n^3)$. Finally, the $k$-means clustering runs in $O(nk^2)$. The dominant term is in practice therefore $O(n^3)$ (we assume we have more points than dimensions), and the impractical running time of Spectral Clustering is due to the need of diagonalization of the Laplacian matrix \cite{li2011time}.

\subsection{Alternative Classical Algorithms} 

Randomness naturally occurs during our quantum spectral clustering, and one may wonder for fair comparison if an alternative classical algorithm can efficiently implement spectral clustering with noisy or sampled methods as well. To circumvent the prohibitive running time of the classical algorithm, several approximations have indeed been proposed on different steps. A recent review of these techniques \cite{tremblay2020approximating} concludes that despite many efforts, methods with provable scalability are found limited or worse in practice, whereas other good empirical methods have no provable guarantees. 

Some methods aim to build the similarity graph using sampling \cite{choromanska2013fast, li2011time,rahimi2008random}. They present limitations \cite{wang2019scalable} and act by sampling partially the input data, which is not the case of our quantum algorithm. Their running time is often proportional to $O(nm)$ or $O(nm^2)$ where $m$ is the number of edges, which is $O(n^2)$ in the worst case. If one of these methods was empirically efficient, we could actually adapt it to our quantum algorithm, by first applying a similar sparsification and then using it as input in our quantum algorithm. Recently such techniques have been done in the quantum setting \cite{apers2019quantum}.

At the next step, it is possible to use Lanczos methods to compute the $k$ lowest eigenvalues and eigenvectors, with a running time of $O(n^2)$ for a fully connected graph. However, they seem to suffer from poor efficiency in practice since they strongly rely on the distribution of the eigenvalues and can require many iterations that would ruin the advantage \cite{bai2000templates}. Note that even with an effective application of this method, our quantum algorithm would still be advantageous. We can also cite the use of power methods \cite{boutsidis2015spectral} to solve clustering using approximated eigenvectors.

Some methods try to improve the clustering step itself, by modifying the $k$-means algorithm \cite{hamerly2015accelerating}. One should compare these methods directly with the quantum $k$-means paper \cite{qmeans}. In most cases, such variations carry over to the quantum case as well. Finally, other attempts use solely preprocessing techniques \cite{yan2009fast} on the initial dataset. Again, one could simply use them before the quantum algorithm to similarly improve its practical efficiency.
\\

\section{Preliminaries in Quantum Information}\label{quantumpreliminaries}

\subsection{Quantum Information}\label{qubitstuto}
We introduce a basic and broad-audience quantum information background necessary for this work. For a more detailed introduction we recommend \cite{NC02}.

\paragraph{Quantum Bits and Quantum Registers:} The bit is the basic unit of classical information. It can be either in state 0 or 1. Similarly a quantum bit, or \emph{qubit}, is a quantum system that can be in state $\ket{0}$, $\ket{1}$ (the \emph{braket} notation $\ket{\cdot}$ is a reminder that the bit considered is a quantum system) or in superposition of both states $\alpha\ket{0}+\beta\ket{1}$ with coefficients $\alpha,\beta \in \mathbb{C}$ such that $|\alpha|^2 + |\beta|^2 = 1$. The \emph{amplitudes} $\alpha$ and $\beta$ are linked to the probabilities of observing either 0 or 1 when \emph{measuring} the qubit, since $P(0)=|\alpha|^2$ and $P(1)=|\beta|^2$. 

Before the measurement, any superposition is possible, which gives quantum information special abilities in terms of computation. With $n$ qubits, the $2^n$ possible binary combinations can exist simultaneously, each with a specific amplitude. For instance we can consider an uniform distribution $\frac{1}{\sqrt{n}}\sum_{i=0}^{2^n-1}\ket{i}$ where $\ket{i}$ represents the $i^{th}$ binary combination (e.g.  $\ket{01\cdots1001}$). Multiple qubits together are often called a \emph{quantum register}. 

In its most general formulation, a quantum state with $n$ qubits can be seen as a vector in a complex Hilbert space of dimension $2^n$. This vector must be normalized under $\ell_2$-norm, to guarantee that the squared amplitudes sum to 1. 

With two quantum states or quantum registers $\ket{p}$ and $\ket{q}$, the whole system is written as a tensor product $\ket{p}\otimes\ket{q}$, often simplified as $\ket{p}\ket{q}$ or $\ket{p,q}$.

\paragraph{Quantum Computation:} To process qubits and therefore quantum registers, we use quantum gates. These gates are \emph{unitary operators} in the Hilbert space as they should map unit-norm vectors to unit-norm vectors. Formally, we can see a quantum gate acting on $n$ qubits as a matrix $U \in \mathbb{C}^{2^n}$ such that $UU^{\dagger}=U^{\dagger}U=I$, where $U^{\dagger}$ is the conjugate transpose of $U$. Some basic single-qubit gates includes the NOT gate 
$\begin{pmatrix}
0 & 1 \\
1 & 0 \\
\end{pmatrix}$ that inverts $\ket{0}$ and $\ket{1}$, or the Hadamard gate 
$\frac{1}{\sqrt{2}}\begin{pmatrix}
1 & 1 \\
1 & -1 \\
\end{pmatrix}$ that maps $\ket{0} \mapsto \frac{1}{\sqrt{2}}(\ket{0}+\ket{1})$ and $\ket{1} \mapsto \frac{1}{\sqrt{2}}(\ket{0}-\ket{1})$, creating the quantum superposition. 

Finally, multiple-qubit gates exist, such as the Controlled-NOT that applies a NOT gate on a target qubit conditioned on the state of a control qubit. 

The main advantage of quantum gates is their ability to be applied to a superposition of inputs. Indeed, given a gate $U$ such that $U\ket{x} \mapsto \ket{f(x)}$, we can apply it to all possible combinations of $x$ at once $U(\frac{1}{C}\sum_{x}\ket{x}) \mapsto \frac{1}{C}\sum_{x}\ket{f(x)}$.

For instance, in the following claim we state some primitive quantum circuits, which we will use in our algorithm. They are basically quantum circuits with a reversible version of the classical ones.
\begin{claim}\label{claimcircuits}
Using quantum circuits, one can perform the following operations in time linear in the number of qubits used to encode the input values :
    \begin{itemize}
    	\item For two integers $i$ and $j$, we can check their equality with the mapping $\ket{i}\ket{j}\ket{0} \mapsto \ket{i}\ket{j}\ket{[i=j]}$. 
    	\item For two real numbers $a>0$ and $b>0$, we can compare them using $\ket{a}\ket{b}\ket{0} \mapsto \ket{a}\ket{b}\ket{[a\leq b]}$. 
    	\item For a real number $a>0$, we can obtain its square $\ket{a}\ket{0} \mapsto \ket{a}\ket{a^2}$.
    \end{itemize}
\end{claim}

\subsection{Quantum Subroutines for Data Loading}\label{subroutinefordata}
Knowing some basic principles of quantum information, the next step is to understand how data can be efficiently encoded using quantum states. While several approaches could exist, we present the most common one called \emph{amplitude encoding}, which leads to interesting and efficient applications. 

Let $x \in \R^{d}$ be a vector with components $(x_1,\cdots,x_{d})$. Then, we define $\ket{x}$ as the quantum state encoding $x$, and given by \begin{equation}
\ket{x} = \frac{1}{\norm{x}}\sum_{j=0}^{d-1}x_j\ket{j}
\end{equation}
We see that the $j^{th}$ component $x_j$ becomes the amplitude of $\ket{j}$, a representation of the $j^{th}$ vector in the standard basis. Each amplitude must be divided by $\norm{x}$ to preserve the unit $\ell_2$-norm of $\ket{x}$. 

Similarly, for a matrix $A \in \R^{n\times d}$ or equivalently for $n$ vectors $A_{i}$ for $i\in [n]$, we can express each row of $A$ as $\ket{A_i} = \frac{1}{\norm{A_i}}\sum_{i=0}^{d-1}A_{ij}\ket{j}$. 

We can now explain an important definition, the ability to have \emph{quantum access} to a matrix. This will be a requirement for many algorithms. In fact one goal of this work is to obtain quantum access to several data related matrices.

\begin{definition}{[Quantum Access to Data]\\}\label{definitionquantumaccess}
We say that we have quantum access to a matrix $A\in \R^{n\times d}$ if there exists a procedure to perform the following mapping, for $i \in [n]$, in time $T$:
\begin{itemize}
    \item $\ket{i}\ket{0} \mapsto \ket{i}\ket{A_i}$
    \item $\ket{0} \mapsto \frac{1}{\norm{A}_F}\sum_{i} \norm{A_i}\ket{i}$
\end{itemize}
\end{definition}

By using appropriate data structures the first mapping can be reduced to the ability to perform a mapping of the form $\ket{i}\ket{j}\ket{0} \mapsto \ket{i}\ket{j}\ket{A_{ij}}$.
The second requirement can be replaced by the ability to perform $\ket{i}\ket{0} \mapsto \ket{i}\ket{\norm{A_i}}$ or to just have the knowledge of each norm. Therefore, using matrices such that all rows $A_i$ have the same norm makes it simpler to obtain quantum access. 

The time or complexity $T$ necessary for the quantum access can be reduced to polylogarithmic dependence in $n$ and $d$ if we consider the access to a Quantum Memory or \emph{QRAM}. The QRAM \cite{KP16} is a specific data structure from which a quantum circuit can allow quantum access to data in time $O(\log{(nd)})$.

We now state important methods for processing quantum information. Their goal is to store some information alternatively in the quantum state's amplitude or in the quantum register as a bitstring. 


\begin{theorem}{[Amplitude Amplification and Estimation \cite{BHMT00}]}\label{AAthm}
Let $U$ be an unitary operator such that $U : \ket{0} \mapsto \sqrt{p}\ket{y}\ket{0} + \sqrt{1-p}\ket{y^{\perp}}\ket{1}$ in time $T$, where $p>0$ is the probability of measuring ``0". Given the ability to implement $U$ and $U^{-1}$, it is possible to obtain with high probability the state $\ket{y}\ket{0}$ using $O(\frac{T}{\sqrt{p}})$ queries to $U$, or to estimate $p$ with relative error $\delta$ using $O(\frac{T}{\delta\sqrt{p}})$ queries to $U$.
\end{theorem}  


\begin{theorem}{[Conditional Rotation]}\label{conditionrotationthm}
Given the quantum state $\ket{a}$, with $a \in [-1,1]$, it is possible to perform $\ket{a}\ket{0} \mapsto \ket{a}(a\ket{0}+\sqrt{1-a}\ket{1})$ with complexity $\widetilde{O}(1)$. 
\end{theorem} 

Using Theorem \ref{conditionrotationthm} followed by Theorem \ref{AAthm}, it then possible to transform the state $\frac{1}{\sqrt{d}}\sum_{j=0}^{d-1} \ket{x_j}$ into $\frac{1}{\norm{x}}\sum_{j=0}^{d-1} x_{j}\ket{x_j}$.

\begin{theorem}{[Tomography of quantum states \cite{KP18}]}\label{tomographythm}
Given a circuit producing in time $T$ a quantum state $\ket{x} = \frac{1}{\norm{x}}\sum_{j=0}^{d-1} x_j\ket{j}$, encoding a vector $x \in \R^d$, there is an algorithm that allows us to output a classical vector $\overline{x}$ with $\ell_2$-norm guarantee $\norm{\overline{x}-x}\leq\delta$ for any $\delta>0$, in time $O(T\times \frac{d\log(d)}{\delta^2})$, with probability at least $1-1/poly(d)$.
\end{theorem}

\subsection{Quantum Subroutines for Linear Algebra}\label{subroutineforlinalg}
In recent years, as the field of quantum machine learning grew, the ``toolkit" for linear algebra algorithms has become important enough to allow the development of many quantum machine learning algorithms. We introduce here the important subroutines for this work, without detailing the circuits or the algorithms. 

\begin{theorem}{[Quantum Singular Value Estimation \cite{KP17}]}\label{svethm}
Given quantum access in time $T$ to a matrix $A\in\R^{m\times n}$ with singular value decomposition $A = \sum_i \lambda_i u_i v_i^T$, there is a quantum algorithm that performs the mapping $\sum_i\alpha_i\ket{u_i}\ket{0} \mapsto \sum_i\alpha_i\ket{u_i}\ket{\overline{\lambda_i}}$ such that for any precision $\epsilon>0$, we have for all singular values $|\overline{\lambda_i} - \lambda_i|\leq \epsilon$, in time $\widetilde{O}(T\times\frac{\mu(A)}{\epsilon})$, with probability at least $1-1/poly(n)$.
\end{theorem} 

\begin{definition}\label{definitionmu}
For a matrix $A$, the parameter $\mu(A)$ is defined by 
$$\mu(A) = \min_{p\in[0,1]}\left(\norm{A}_F, \sqrt{s_{2p}(A)s_{2(1-p)}(A^T)}\right)$$
 where $s_p(A) = \max_{i}(\norm{A_i}_p^p)$. 
\end{definition}

For dense matrices, $\mu(A)$ can be taken to be the ratio Frobenius norm over spectral norm of $A$. In some sense, it replaces the explicit dependence on the matrix dimension. For sparse matrices, $\mu(A)$ can be seen as the sparsity.

The next theorem allows us to compute the distance between vectors encoded as quantum states. It is a fundamental routine used for instance to construct the similarity graph.
\begin{theorem}{[Quantum Distance Estimation \cite{wiebe2014quantum,qmeans}]}\label{distancethm}
Given quantum access in time $T$ to two matrices $U$ and $V$ with rows $u_i$ and $v_j$ of dimension $d$, there is a quantum algorithm that, for any pair $(i,j)$, performs the following mapping  $\ket{i}\ket{j}\ket{0} \mapsto \ket{i}\ket{j}\ket{\overline{d^2(u_i,v_j)}}$, estimating the Euclidean distance between $u_i$ and $v_j$ with precision $|\overline{d^2(u_i,v_j)} - d^2(u_i,v_j)| \leq \epsilon$ for any $\epsilon>0$. The algorithm has a running time given by $\widetilde{O}(T\eta/\epsilon)$, where $\eta = \max_{ij}(\norm{u_i}\norm{v_j})$, assuming that $\min_{i}(\norm{u_i}) = \min_{i}(\norm{v_i}) = 1$.
\end{theorem} 

In the following, we will present the result of \cite{qmeans}, a quantum analogue of the $k$-means clustering that can provide a speedup. The $k$-means algorithm finds, among unlabeled data points, the center (or centroid) of each cluster forming the dataset. It proceeds by iteratively estimating the distance between each data point and each centroid to update the nearest centroid for each point, then calculate the new centroid for each cluster, until convergence is reached.

We now introduce new parameters used in the quantum version. The parameter $\delta > 0$ is the precision parameter used in both the estimation of the distances between vectors and in the estimation of the position of the cluster's centroids. $V$ is the input data matrix, $\kappa(V)$ is its condition number and $\mu(V)$ is defined above (Definition \ref{definitionmu}). We also need the following definition :

\begin{definition}\label{etadefinition}
For a matrix $V \in \R^{n\times d}$, its parameter $\eta(V)$ is defined as $\frac{\max_i(\norm{v_i}^2)}{\min_{i}(\norm{v_i}^2)}$, or as $\max_i(\norm{v_i}^2)$ assuming $\min_{i}(\norm{v_i})=1$.
\end{definition}

We finally recall the \emph{well-clusterability} assumption from \cite{qmeans}, which assures a specific structure in the data to be clustered, ensuring efficiency for both classical and quantum $k$-means clustering. In spirit, this assumption guarantees that almost all data points can be easily assigned to a cluster and that each cluster's diameter is sufficiently small (intra-cluster distance) compared to the distance between each other (inter-cluster distance). See \cite{qmeans} for the detailed description.

\begin{definition}\label{wellclusterabilitydef}
A data matrix $V \in \R^{n \times d}$ with rows $v_{i} \in \R^{d}, i \in [N]$, is said to be well-clusterable if there exist constants $\xi, \beta>0$, $\lambda \in [0,1]$, $\eta \leq 1$, and cluster centroids $c_i$ for $i\in [k]$ such that the following hold:\\
- (separation of cluster centroids): $d(c_i, c_j) \geq \xi \quad \forall i,j \in[k]$ \\
- (proximity to cluster centroid): At least $\lambda n$ points $v_i$ in the dataset satisfy 
$d(v_i, c_{l(v_i)}) \leq \beta$ where $c_{l(v_i)}$ is the centroid 
    nearest to $v_{i}$. \\
- (Intra-cluster smaller than inter-cluster square distances): 
 The following inequality is satisfied $4\sqrt{\eta} \sqrt{ \lambda \beta^{2} + (1-\lambda) 4\eta} \leq \xi^{2} - 2\sqrt{\eta} \beta.$
\end{definition}

\begin{theorem}{[Quantum k-means clustering \cite{qmeans}]\\}\label{qmeansthm}
Given quantum access in time $T_V$ to a dataset $V\in\R^{n\times d}$, there is a quantum algorithm that outputs with high probability $k$ centroids $c_1,\cdots,c_k$ that are consistent with the output of the $k$-means algorithm with noise $\delta>0$, in time 
$$
\widetilde{O}\left( T_V  T_{qmeans} \right) = 
\widetilde{O}\left(T_V \left( kd\frac{\eta(V)}{\delta^2}\kappa(V)(\mu(V) + k\frac{\eta(V)}{\delta}) + k^2\frac{\eta(V)^{1.5}}{\delta^2}\kappa(V)\mu(V)\right)\right)
$$
per iteration. If the dataset $V$ is well-clusterable, the running time simplifies to 
$$
\widetilde{O}\left( T_V  T_{qmeans} \right) = 
\widetilde{O}\left(T_V\left( k^2d\frac{\eta(V)^{2.5}}{\delta^3} + k^{2.5}\frac{\eta(V)^{2}}{\delta^3}\right)\right)
$$

\end{theorem}

Note that the dependence in $n$ is hidden in the time $T$ to load the data as well as the other data-driven parameters that may depend on $n$. \\

\section{Quantum Algorithm for Graph Based Machine Learning and Spectral Clustering}\label{quantumspectralclusteringsection}

In this part, we will detail the quantum algorithm that performs Spectral Clustering with similar guarantees and more efficient running time compared to its classical analogue. We start by presenting all the theorems that will help us construct the Quantum Spectral Clustering algorithm, then in the following subsections we provide details and proofs. 

We first state the theorem that allows us to compute the edge's value of the data adjacency graph:

\begin{theorem}{[Quantum Algorithm for Data Point Similarity]\\}\label{quantumaccestosimilarity}
Given quantum access to the data matrix $S \in \R^{n\times d}$ in time $T_S$ and two indices $p,q \in [n]^2$, we can obtain the following mapping: $\ket{p}\ket{q}\ket{0} \mapsto \ket{p}\ket{q}\ket{a_{pq}}$ in time $O(T_S \times \frac{\eta(S)}{\epsilon_{dist}})$. The elements $a_{pq}$ correspond to the edge's values of the data adjacency graph, using the rule of construction based on a threshold distance. $\eta(S)$ is a data parameter defined in Definition \ref{etadefinition}, $\epsilon_{dist}>0$ is the precision parameter in the estimation of the distance between input points. 
\end{theorem}

Using the previous theorem we can have quantum access to the normalized Incidence matrix.

\begin{theorem}{[Quantum access to the Normalized Incidence Matrix]\\}\label{quantumaccesstoB}
Given quantum access to the data matrix $S \in \R^{n\times d}$ in time $T_S$ we can have quantum access to the normalized incidence matrix $\mathcal{B} \in \R^{n\times \frac{n(n-1)}{2}}$ in time $T_{\mathcal{B}} = \widetilde{O}(T_S\times \frac{\eta(S)}{\epsilon_{dist}} \times \frac{1}{\epsilon_{B}})$, where $\eta(S)$ is defined in Definition \ref{etadefinition}, $\epsilon_{dist} >0$ is the precision of distance estimation between vectors, and $\epsilon_{B}$ is the substitute of the zeros in $B$.
\end{theorem}

Using the previous theorem we can finally have quantum access to the projected Laplacian matrix.

\begin{theorem}{[Quantum access to projected Laplacian matrix]\\}\label{quantumaccesstoL}
Given quantum access to the normalized incidence matrix $\mathcal{B} \in R^{n\times \frac{n(n-1)}{2}}$ in time $T_{\mathcal{B}}$, we can have quantum access to $\widetilde{\mathcal{L}}^{(k)} \in R^{n\times k}$, the Laplacian matrix projected onto its $k$ lowest eigenvalues, in time $T_{\widetilde{\mathcal{L}}^{(k)}} = \widetilde{O}(T_{\mathcal{B}}  \frac{\mu(\mathcal{B})\kappa(\widetilde{\mathcal{L}}^{(k)})}{\epsilon_{\lambda}})$, where $\epsilon_\lambda$ is the precision parameter for estimating the eigenvalues of $\mathcal{L}$, $\mu(\mathcal{B})$ is a data parameter defined in Definition \ref{definitionmu}, and $\kappa(\widetilde{\mathcal{L}}^{(k)})$ is the condition number of $\widetilde{\mathcal{L}}^{(k)}$.
\end{theorem}

The Quantum Spectral Clustering algorithm consists then of applying the quantum $k$-means algorithm (Theorem \ref{qmeansthm}) using the fact that we have quantum access to the projected Laplacian matrix.

The main algorithm can thus be summarized in the following theorem. 



\begin{theorem}
{[Quantum Spectral Clustering]\\}
Given quantum access to a data matrix $S \in \R^{n\times d}$ in time $T_S$, there is a quantum algorithm that with high probability outputs $k$ centroids in the Laplacian spectral space, in time 
$$
\widetilde{O}
    \left( T_S
    \frac{\eta(S)}{\epsilon_{dist}\epsilon_B}     
    \frac{\mu(\mathcal{B})\kappa(\widetilde{\mathcal{L}}^{(k)})}{\epsilon_{\lambda}}
    T_{qmeans}
    \right)
$$
where $T_{qmeans}$ is the multiplicative factor in the running time of the quantum clustering algorithm (Theorem  \ref{qmeansthm}) on the vectors projected onto the Laplacian spectral space. In the case where the vectors projected onto the spectral space are \emph{well-clusterable} (see Definition \ref{wellclusterabilitydef}), the running time becomes
$$
\widetilde{O}
    \left( T_S 
    \frac{\eta(S)}{\epsilon_{dist}\epsilon_B}     
    \frac{\mu(\mathcal{B})\kappa(\widetilde{\mathcal{L}}^{(k)})}{\epsilon_{\lambda}}     
    \frac{k^3\eta({\widetilde{\mathcal{L}}^{(k)}})^{2.5}}{\delta^3}
    \right)
$$
In the above formulas, $\mathcal{B}$ refers to the normalized incidence matrix of the data, and $\widetilde{\mathcal{L}}^{(k)}$ to the projection of the Laplacian matrix. $\epsilon_{dist}$, $\epsilon_{B}$, $\epsilon_\lambda$ and $\delta$ are error or precision parameters. $\eta(S)$, $\eta(\widetilde{\mathcal{L}}^{(k)})$, and $\mu(\mathcal{B})$ are data parameters defined in Definition \ref{etadefinition} and Definition \ref{definitionmu}, and $\kappa(\widetilde{\mathcal{L}}^{(k)})$ is the condition number of $\widetilde{\mathcal{L}}^{(k)}$.
\end{theorem}

\subsection{Computing the similarity between two nodes}\label{similarity_section}

\begin{figure}[t!]
\centering
   \includegraphics[width=460px]{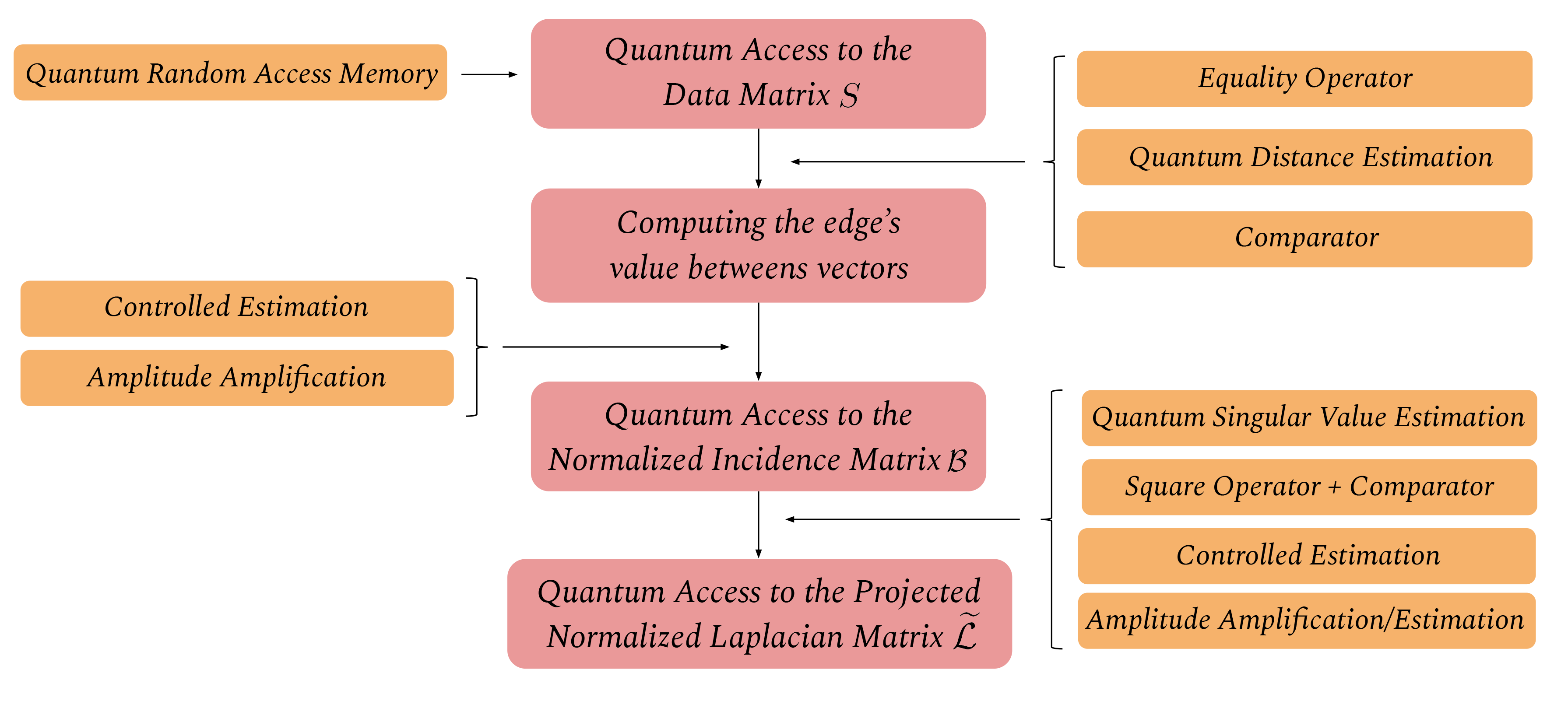}
   \label{fig_diagram} 
\caption{Diagram for Sections \ref{similarity_section} to \ref{SVEsection}.}
\label{diagram}
\end{figure}

We now detail the algorithm for Theorem \ref{quantumaccestosimilarity}, which allows us to compute the elements $a_{pq}$ of the Adjacency Matrix, corresponding to a pair of input vectors $s_p$ and $s_q$. This will be used as a subroutine in the algorithm that gives quantum access to the normalized incidence matrix (see Fig. \ref{diagram}).

For any two quantum states corresponding to the indices $\ket{p}$ and $\ket{q}$, we can use Theorem \ref{distancethm} to obtain $\ket{p}\ket{q}\ket{\overline{d(s_p,s_q)^2}}$. The square distance obtained is approximated with a precision $\epsilon_{dist} > 0$ such that $|\overline{d(s_p,s_q)^2} - d(s_p,s_q)^2|\leq \epsilon_{dist}$. 

These distances are then converted into the edge values $a_{pq} \in \{0,1\}$ using our modified graph construction rule from Section \ref{notationsanddefinition}. For doing so, we can use the comparison operator (see Claim \ref{claimcircuits}) to check if $\overline{d(s_p,s_q)^2}$ is smaller than the desired threshold below which we consider that two nodes are connected. Therefore we can easily obtain $\ket{p}\ket{q}\ket{a_{pq}}$.

Note that it was necessary to use this particular definition of the Adjacency Matrix in order to be able to perform this operation efficiently quantumly. Other definitions include the direct use of the distance, or sometimes a mapping of the distance using diverse kernels.

Finally, note that using this theorem, it is possible to have quantum access to a normalized version of the Adjacency Matrix, which could be useful in many applications. In our case, however, it is not clear how to use it to obtain quantum access to the normalized Laplacian, because it would require the Degree matrix and the norms $\norm{a_i}$, which are not accessible efficiently. Therefore, we chose to work with the \emph{normalized} Incidence matrix.

\subsection{Quantum access to the Normalized Incidence Matrix $\mathcal{B}$}\label{NormalizedB_section}

To obtain quantum access to $\mathcal{B}$ (see Definition \ref{definitionquantumaccess}), we start with a quantum state encoding one node's index $i \in [n]$, along with the superposition of all $\frac{n(n-1)}{2}$ possible edges encoded as pairs $(p,q) \in [n]^2$ with $p<q$. We use the following quantum state, whose preparation requires $O{(\log(n))}$ qubits and a simple quantum circuit:
\begin{equation}
\frac{\sqrt{2}}{\sqrt{n(n-1)}}
\ket{i}
\sum_{p<q}\ket{p}\ket{q}
\end{equation}

From this state, we first determine which edges are incident using two ancillary qubits as flags, using the equality operator (see Section \ref{qubitstuto}). This allows us to separate the cases that appear in Eq. (\ref{Bdefinition}). For simplicity, we will use the notation $\ket{p}\ket{q} = \ket{p,q}$:

\begin{equation}\label{afterdist}
    \frac{\sqrt{2}}{\sqrt{n(n-1)}}
    \ket{i}
    \left(
    \sum_{\substack{p<q \\ i=p}}\ket{p,q}\ket{11}
    +
    \sum_{\substack{p<q \\ i=q}}\ket{p,q}\ket{10}
    +
    \sum_{\substack{p<q \\ i\notin \{p,q\}}}\ket{p,q}\ket{00}
    \right)
\end{equation}

We then use another register to write the values of $B$, the unnormalized incidence matrix, given by Eq. (\ref{Bdefinition}). For the flagged edges ($i=p$ or $i=q$), using a controlled version of Theorem \ref{quantumaccestosimilarity}, we obtain the superposition of the similarity between all input points and point $i$, in a quantum register. For the other edges ($i\notin \{p,q\}$), instead of simply writing 0 as in Eq. (\ref{Bdefinition}) we modify the zero elements of the matrix to a value $\epsilon_B > 0$ in order to retain the efficiency of the quantum algorithm in the next step. The running time for this step is $O(T_S \times \eta(S)/\epsilon_{dist})$, where $\eta(S)$ is a data parameter defined in Definition \ref{etadefinition}, and $\epsilon_{dist}>0$ is the precision parameter in the estimation of $d^2(s_i,s_j)$. Finally $T_S$ is the time to have quantum access to the input points $s_1,\cdots,s_n$, which becomes $T_S = O(\log(nd))$ if we assume QRAM access.

We obtain the following state:

\begin{equation}
    \frac{\sqrt{2}}{\sqrt{n(n-1)}}
    \ket{i}
    \left(
    \sum_{\substack{p<q \\ i=p}}\ket{p,q}\ket{11}\ket{a_{pq}}
    +
    \sum_{\substack{p<q \\ i=q}}\ket{p,q}\ket{10}\ket{a_{pq}}
    +
    \sum_{\substack{p<q \\ i\notin \{p,q\}}}\ket{p,q}\ket{00}\ket{\epsilon_B}
    \right)
\end{equation}

Which, by Eq. (\ref{Bdefinition}) and after uncomputing the flags would be equal to
\begin{equation}
   \frac{\sqrt{2}}{\sqrt{n(n-1)}}
    \ket{i}
    \sum_{p<q}\ket{p,q}\ket{B_{i,(p,q)}}
\end{equation}
From this state, we use a conditional rotation (see Theorem \ref{conditionrotationthm}) to encode, in superposition, the values of the $B$ into the amplitude of a new qubit, and after uncomputing the values of the matrix $B$ from the registers, we have the state:  

\begin{equation}
    \frac{\sqrt{2}}{\sqrt{n(n-1)}}
    \ket{i}
    \left(
    \sum_{p<q}B_{i,(p,q)} \ket{p,q} \ket{0}+ \sum_{p<q}\sqrt{1-B_{i,(p,q)}^2} \ket{p,q} \ket{1}
    \right)
\end{equation}

Finally, an amplitude amplification (see Theorem \ref{AAthm}) is performed to select the $\ket{0}$ part of the state, and obtain a valid quantum encoding of the vector $B_i$. This requires us to repeat the previous steps $O(1/\sqrt{P_i(0)})$ times where $P_i(0)$ is the probability of reading $``0"$ in the last register. Since all elements of the matrix $B$ have norm at least $\epsilon_B$ we therefore have that $O(1/\sqrt{P_i(0)})$ is at least $O(1/\epsilon_B)$. We finally obtain with high probability the state
$
    \ket{i}
    \frac{1}{\norm{B_i}}\sum_{p<q}B_{i,(p,q)}\ket{p,q}
    = \ket{i}\ket{\mathcal{B}_i}
$.

Recall from Section \ref{notationsanddefinition} that $\mathcal{B}_{i,(p,q)} = \frac{B_{i,(p,q)}}{\norm{B_i}}$ and that $\norm{B_i} = \norm{a_i}$. In addition, we have access to the norm of each row of $\mathcal{B}$ since by definition they are all equal to $1$. We can therefore conclude that we have quantum access to $\mathcal{B}$, the normalized incidence matrix, according to Definition (\ref{definitionquantumaccess}). The global running time to have quantum access to $\mathcal{B}$ is given by $O(T_S \frac{\eta(S)}{\epsilon_{dist}} \frac{1}{\epsilon_B}))$. This proves Theorem \ref{quantumaccesstoB}.

\subsection{Quantum Access to the Projected Normalized Laplacian Matrix $\widetilde{\mathcal{L}}^{(k)}$}\label{SVEsection}

With the quantum access to the normalized incidence matrix $\mathcal{B}$ in time $T_{\mathcal{B}}$, we will use the fact that the $i^{th}$ row of $\mathcal{L}=\mathcal{B}\mathcal{B}^T$ can be written as $\mathcal{L}_i = \mathcal{L}\cdot e_i$ where $e_i$ represents the $i^{th}$ vector of the standard basis, for which the corresponding quantum state is simply $\ket{i}$. We also use the fact that this state can be naturally expressed as $\ket{i} = \sum_j\sigma_{ij}\ket{u_j}$ in the basis made of the left singular vectors $u_j$ of $\mathcal{B}$, with unknown coefficients $\sigma_{ij}$, such that $\sum_j\sigma_{ij}^2=1$.

On this initial state $\ket{i}$ we apply the SVE algorithm\footnote{Since SVE must be applied to a square matrix, we use in fact $\begin{pmatrix}0 & \mathcal{B}\\\mathcal{B}^T & 0\end{pmatrix}$; see \cite{KP16}.} 
(Theorem \ref{svethm}) to estimate the singular values of $\mathcal{B}$ in superposition and obtain $\sum_j\sigma_{ij}\ket{u_j}\ket{\lambda^{\mathcal{B}}_j}$. The running time for this step is $O(T_\mathcal{B} \times \mu(\mathcal{B})/\epsilon_{\lambda})$, where $\mu(\mathcal{B})$ is a data parameter defined in Definition \ref{definitionmu}, and $\epsilon_{\lambda}>0$ is the desired precision in the estimation of the singular values. We then square these values to obtain the state $\sum_j\sigma_{ij}\ket{u_j}\ket{\lambda_j}$, with the eigenvalues of $\mathcal{L}$. Note that $\mu(\mathcal{B})$ is upper bounded by $n$ \cite{KP16}.

At this point, we can prepare the projection on the $k$ lowest eigenvectors of $\mathcal{L}$ by first separating the eigenvalues lower that a threshold $\nu > 0$ with an ancillary qubit, such that the $k$ lowest eigenvalues are flagged by $``0"$ :
$
    \sum_{\substack{j | \lambda_j \leq \nu}}\sigma_{ij}\ket{u_j}\ket{\lambda_j}\ket{0}
    + 
    \sum_{\substack{j | \lambda_j > \nu}}\sigma_{ij}\ket{u_j}\ket{\lambda_j}\ket{1}
$.

If the flag qubit is $``0"$, we perform a conditional rotation on the eigenvalue register using Theorem \ref{conditionrotationthm} in a controlled fashion. For this we introduce again an ancillary qubit. Since the amplitude of the rotation must be inferior to 1, we first divide each eigenvalue by $\nu$, which is higher than the largest eigenvalue of $\widetilde{\mathcal{L}}^{(k)}$ :  
\begin{equation}\label{LaplacianCR}
    \sum_{\substack{j | \lambda_j \leq \nu}}\sigma_{ij}\ket{u_j}\ket{\lambda_j}\ket{0}
    \left(\frac{\lambda_{j}}{\nu}\ket{0} + \sqrt{1-\frac{\lambda^2_j}{\nu^2}}\ket{1}\right)
    +
    \sum_{\substack{j | \lambda_j > \nu}}\sigma_{ij}\ket{u_j}\ket{\lambda_j}\ket{1}\ket{0}
\end{equation}

From this quantum state, we will show how to obtain quantum access to $\ket{\widetilde{\mathcal{L}}^{(k)}}$ using amplitude estimation and amplitude amplification on the $``00"$ value on the last two registers. We recall that quantum access is guaranteed if we can recover, for each row $\widetilde{\mathcal{L}}^{(k)}_i$, its norm and the corresponding quantum state.\\
Let  $P_i(00)$ be the probability of reading $``00"$ on the last two registers. It is easy to show that $ P_i(00) = \frac{1}{\nu^2}\sum_{\substack{j | \lambda_j \leq \nu}}\sigma_{ij}^2\lambda_{j}^2$. 
We will prove that $\norm{\widetilde{\mathcal{L}}^{(k)}_i}= \nu\sqrt{P_i(00)}$, and that amplifying the state will yield to $\ket{\widetilde{\mathcal{L}}^{(k)}_i}$. \\
The normalized Laplacian matrix can be written in its eigenbasis using the outer product $\mathcal{L} = \sum_j \lambda_j \ket{u_j}\bra{u_j}$. 
Similarly, recall that we used $\sum_j \sigma_{ij} \ket{u_j}$ to encode $e_i$, the $i^{th}$ vector in the standard basis. Therefore, each row $\mathcal{L}_i = \mathcal{L} \cdot e_i$ can be written as $\ket{\mathcal{L}_i} = \sum_j \lambda_j \ket{u_j}\bra{u_j} \cdot \sum_j \sigma_{ij} \ket{u_j} = \sum_j \sigma_{ij}\lambda_j\ket{u_j}$. We thus have a formula for the norm of the rows $\norm{\mathcal{L}_i}=\sqrt{\sum_j \sigma_{ij}^2\lambda_j^2}$. The same idea holds for $\widetilde{\mathcal{L}}^{(k)}$, the projection of the normalized Laplacian, and we obtain $\norm{\widetilde{\mathcal{L}}^{(k)}_i}=\sqrt{\sum_{j=1}^{k}\sigma_{ij}^2\lambda_j^2} = \nu\sqrt{P_i(00)}$.\\
Therefore, using amplitude estimation (Theorem \ref{AAthm}), we can estimate $P(00)$ and have access to the norms of $\widetilde{\mathcal{L}}^{(k)}$ to a multiplicative constant. Similarly, using amplitude amplification (Theorem \ref{AAthm} also), we can amplify the $\ket{00}$ state and obtain the state $\sum_{\lambda_j \leq \nu}\sigma_{ij}\lambda_j\ket{u_j} = \ket{\widetilde{\mathcal{L}}^{(k)}_i}$.\\
The number of iterations for amplitude estimation and amplification is $\widetilde{O}(1/\sqrt{P_i(00)})$. Let $\lambda_k$ be the $k^{th}$ eigenvalue of $\mathcal{L}$ and therefore the largest eigenvalue of $\widetilde{\mathcal{L}}^{(k)}$. By correctly choosing the threshold $\nu$ close to the largest eigenvalue, for instance $\nu \leq \gamma \lambda_k$ with $\gamma>1$ a small positive constance, we can write:
\begin{equation}
P_i(00) =\sum_{j | \lambda_j \leq \nu}\sigma_{ij}^2 \frac{\lambda_{j}^2}{\nu^2} 
\geq 
\sum_{j | \lambda_j \leq \lambda_k}\sigma_{ij}^2 \frac{\lambda_{min}^2}{\gamma^2 \lambda_k^2}
= \frac{1}{\gamma^2} \frac{\lambda_{min}^2}{\lambda_k^2} 
= \frac{1}{\gamma^2} \frac{1}{\kappa(\widetilde{\mathcal{L}}^{(k)})^2} 
= O\left(1 /\kappa(\widetilde{\mathcal{L}}^{(k)})^2\right)
\end{equation}

Therefore the number of iterations for this step can be upper bounded by $\widetilde{O}(\kappa(\widetilde{\mathcal{L}}^{(k)}))$. Overall, the running time is given by $O(T_{\mathcal{B}} \times \frac{\mu(\mathcal{B})}{\epsilon_{\lambda}}\times 
\kappa(\widetilde{\mathcal{L}}^{(k)}))$. This concludes the algorithm that gives quantum access to the projected Laplacian of a graph and proves Theorem \ref{quantumaccesstoL}.  

\subsection{Quantum Clustering in the Spectral Space}
Having quantum access to the projected normalized Laplacian $\widetilde{\mathcal{L}}^{(k)}$, we possess all requirements to apply the quantum $k$-means algorithm, or $q$-means (Theorem \ref{qmeansthm}). We initialize $k$ centroids at random or using $q$-means++, equivalent to the $k$-means++ initialization. The $q$-means algorithm first consists of constructing a state where all distances between rows $\widetilde{\mathcal{L}}^{(k)}_i$ and the current centroids are computed in parallel. The rest of the algorithm consists of finding the label for each row, and updating the centroids as the average of the vectors composing the cluster. After several iterations, the centroids should have converged and can be retrieved. 

Several approximations are necessary during the steps, hence the presence of precision parameter $\delta>0$. It expresses the approximation error committed during the distance estimation, and during the classical description of the new centroids at each step. Therefore, to be more specific, $q$-means is a noisy version of $k$-means. 

In high level, denoting with $T_{\widetilde{\mathcal{L}}^{(k)}}$ the time to have quantum access to $\widetilde{\mathcal{L}}^{(k)} \in \R^{n\times k}$, which plays the role of the initial state for the clustering algorithm,  and $T_{qmeans}$ the remaining multiplicative factor in the running time of the quantum clustering algorithm (see Theorem \ref{qmeansthm}), the overall running time of our algorithm is given by 
\begin{equation}
    T_{\widetilde{\mathcal{L}}^{(k)}} T_{qmeans}
\end{equation}
To go further, we can assume that the vectors are effectively made of well separated clusters. Indeed, it should be the case in the Spectral Clustering method, once projected onto the spectral space (see Section \ref{NumericalSimulations} for numerical simulations). This \emph{well-clusterability} assumption (See Definition \ref{wellclusterabilitydef}) ensures the classical spectral clustering to classify the data accurately, and the quantum algorithm to work efficiently. Indeed, the running time of the $q$-means algorithm is bounded with better guarantees in this case \cite{qmeans}. With this assumption, and with input dimension $k$ for the spectral space of $\widetilde{\mathcal{L}}^{(k)}$, the running time to update the $k$ centroids in the case of \emph{well-clusterable} data is given by
\begin{equation}\label{runningtimeqmeansonL}
\widetilde{O}\left(T_{\widetilde{\mathcal{L}}^{(k)}}  \frac{k^3\eta({\widetilde{\mathcal{L}}^{(k)}})^{2.5}}{\delta^3}\right)
\end{equation}
Again, it is important to note that our algorithm could work without this \emph{well-clusterability} assumption, the only difference would be a different bound on the theoretical runtime (see Theorem \ref{qmeansthm} for the general formulation).


\subsection{Running Time}\label{conclusionrunningtime}
Using the running times obtained in Theorems \ref{quantumaccesstoB}
and \ref{quantumaccesstoL}, as well as result (\ref{runningtimeqmeansonL}) for the \emph{well-clusterable} case, we can conclude that our quantum algorithm for Spectral Clustering has the following running time :

\begin{equation}\label{runtime}
    \widetilde{O}
    \left( T_S     
    \frac{\eta(S)}{\epsilon_{dist}\epsilon_B}     
    \frac{\mu(\mathcal{B})\kappa(\widetilde{\mathcal{L}}^{(k)})}{\epsilon_{\lambda}}     
    \frac{k^3\eta({\widetilde{\mathcal{L}}^{(k)}})^{2.5}}{\delta^3}
    \right)
\end{equation}

\noindent
$T_S$ is the time to have \emph{quantum access} (see Definition \ref{definitionquantumaccess}) to the input vectors $S \in \R^{n\times d}$ which becomes $O(polylog(n,d))$ if we assume access to the QRAM. $k$ is the number of clusters. $\kappa(\widetilde{\mathcal{L}}^{(k)})$ is the condition number of $\widetilde{\mathcal{L}}^{(k)}$. $\mu(\mathcal{B})$, $\eta(S)$, and $\eta({\widetilde{\mathcal{L}}^{(k)}})$ are data parameters defined in Section \ref{subroutineforlinalg}. $\epsilon_B$ is the chosen minimum value in the incidence matrix. $\epsilon_{dist}$ is the precision in the distance calculation between input points. $\delta$ is the precision of the $q$-means algorithm.

\section{Numerical Simulations}\label{NumericalSimulations}

The quantum algorithm performs the same steps as the classical one while introducing noise or randomness along the way. We present numerical simulations on simple synthetic datasets made of two concentric circles, as in the original work on spectral clustering \cite{ng2002spectral}, in order to benchmark the quality of the quantum algorithm. In this way, we see how the quantum effects are impacting the graph and the spectral space, and we obtain numerical estimates on the quantum running time. 

\begin{figure}[h]
\begin{subfigure}[b]{\linewidth}
\centering
   \includegraphics[width=450px]{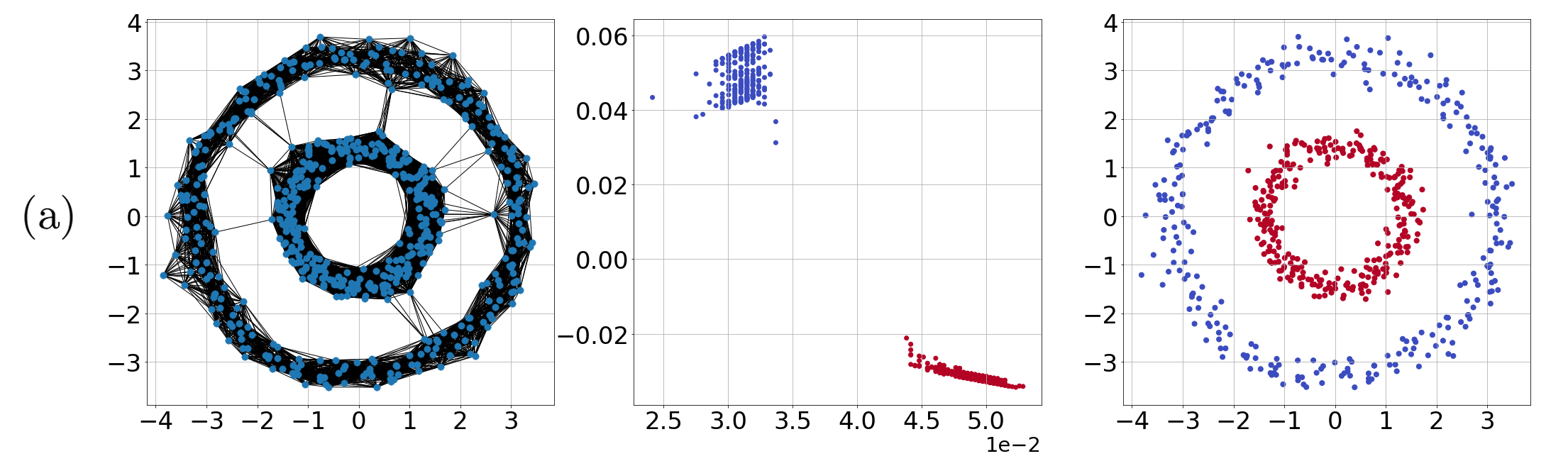}
\end{subfigure}
\begin{subfigure}[b]{\linewidth}
\centering
   \includegraphics[width=450px]{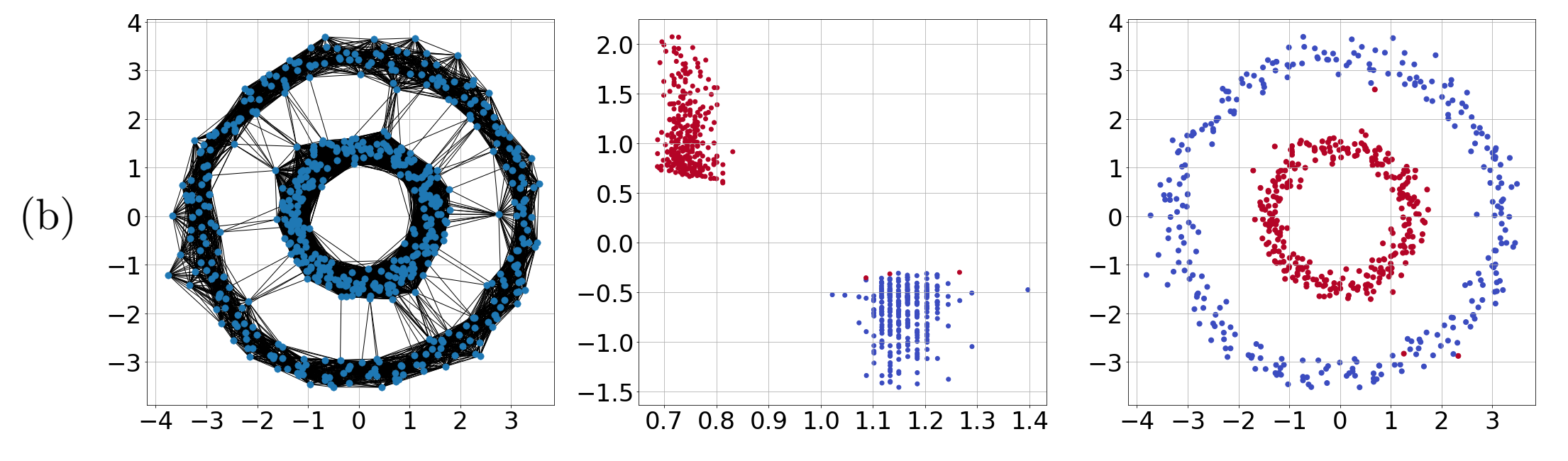}
\end{subfigure}
\caption{(a) Classical and (b) quantum spectral clustering algorithms on two non linearly separable sets of 600 input vectors. The Laplacian is derived from the adjacency graph \emph{(left)}, itself constructed from the data points. The clustering is shown in both spectral \emph{(center)} and input \emph{(right)} space domains. Three points were misclassified in the quantum case.}
\label{figurecircles}
\end{figure}

These simulations are made with a classical computer that simulates the quantum steps and introduces equivalent noise and randomness. It would be very interesting to perform the same experiments using a real quantum computer, alas such computers are not yet available. While simulating the quantum steps, the computation becomes very soon impractical, in fact the simulations we present already take several hours to execute, and thus we leave numerical simulations on larger datasets as future work. Our goal was to design a quantum spectral clustering algorithm with a rigorous theoretical analysis of its running time and provide initial evidence of its practical efficiency and accuracy on a canonical dataset. 

\begin{figure}[h!]
\begin{floatrow}
\ffigbox{%
 \includegraphics[width=220px]{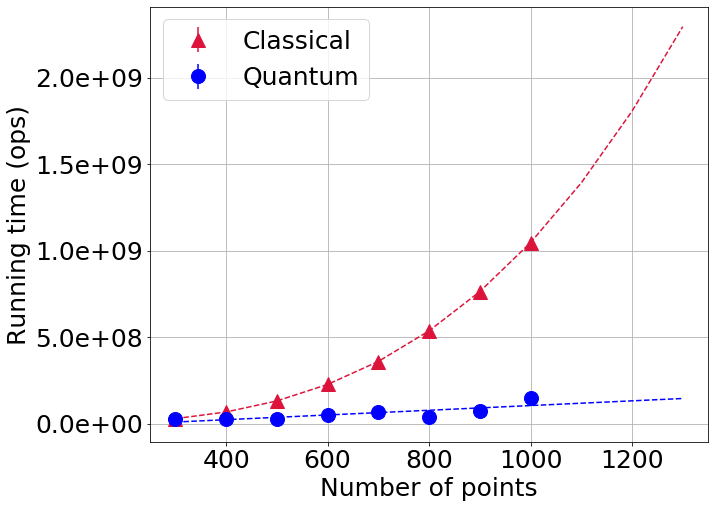} 
}{%
\caption{Running times for quantum and classical spectral clustering. Error bars are present.} 
\label{runningtimes}
}
\capbtabbox{%
  \begin{tabular}{ccc}
    \toprule
    \multicolumn{1}{c}{\multirow{2}{*}{\shortstack[l]{Numbers \\ of points}}} & \multicolumn{2}{c}{Accuracy}                   \\
    \cmidrule(r){2-3}
         &  Classical    & Quantum \\
    \midrule
300 & $100\%$ & $99.6 \% \pm 1.05 \%$  \\
400 & $100\%$ & $99.5 \% \pm 0.83 \%$  \\
500 & $100\%$ & $99.3 \% \pm 1.52 \%$  \\
600 & $100\%$ & $99.9 \% \pm 0.31 \%$  \\
700 & $100\%$ & $100.0 \% \pm 0.0 \%$  \\
800 & $100\%$ & $99.6 \% \pm 0.62 \%$  \\
900 & $100\%$ & $99.6 \% \pm 0.68 \%$  \\
1000 & $100\%$ & $99.9 \% \pm 0.19 \%$  \\
    \bottomrule
  \end{tabular}
}{%
  \caption{Our quantum algorithm finds the clusters with a similar high accuracy.}%
  \label{accuracytable}
}
\end{floatrow}
\end{figure}

The precision parameters used in the quantum case were: $\epsilon_{dist} = 0.1$ for the creation of edges, $\epsilon_{B} = 0.1$ for the creation of the incidence matrix, $\epsilon_{\lambda} = 0.9$ during the estimation of the eigenvalues of $\mathcal{L}$ and finally the precision parameter in $q$-means $\delta = 0.9$. The quantum algorithm was able to classify the two sets with high accuracy (Table \ref{accuracytable}). The clustering was simulated for 300 to 1000 points, repeated 10 times each. In Fig. \ref{figurecircles} we observe clearly the impact of the quantum effects in the graph, where the edges are different, and in the spectral space, where the clusters are more spread out. It is surprising that the quantum algorithm is already faster for small values of $n$, below 1000 points (Fig. \ref{runningtimes}). This difference should substantially increase as $n$ grows. Indeed, compared to the classical algorithm used in practice which scales as $O(n^3)$, the quantum running time is advantageous as its scaling appears to be linear in $n$. In fact, the scaling comes from the factor $\mu(B)$ which is upper bounded by $n$ (see Section \ref{SVEsection}). Note, of course, that both for the classical and quantum running time we used as a proxy the order of steps in the theoretical analysis, disregarding questions of clock time or error correction. Our results show more than anything that it is certainly worth pursuing quantum algorithms for spectral clustering and other graph-based machine learning algorithms, since at least at a first level of comparison they can offer considerable advantages compared to the classical algorithms. It remains an open question to see when and if quantum hardware can become good enough to offer such advantages in practice.

\section{Conclusions}\label{conclusions}
In this work, we described a quantum machine learning algorithm, inspired by the classical Spectral Clustering algorithm. While introducing a number of modifications, approximations, and randomness in the process, the quantum algorithm can still perform clustering tasks with similar very good accuracy, and with a more efficient running time thanks to a weaker dependence on the number of input points: at most linear in our preliminary experiments. This could allow quantum Spectral Clustering to be applied on datasets that are now considered infeasible in practice. Our quantum algorithm is end-to-end, from classical input to classical output, and could pave the way to other graph-based methods in machine learning and optimization, for example using our methods for obtaining access to the normalized Adjacency, Incident, and Laplacian matrices.

\paragraph{Acknowledgments} This work was supported by ANR quBIC, quData, and QuantERA project QuantAlgo. The authors thank Ronald de Wolf, Simon Apers, Batiste Le Bars, and YaoChong Li for their insights, comments, and fruitful discussions.

\paragraph{Data Availability}
The synthetic dataset used in the current study is available in Scikit Learn \cite{scikit-learn}, \url{https://scikit-learn.org/stable/modules/generated/sklearn.datasets.make_circles.html#sklearn.datasets.make_circles}.

\paragraph{Author Contributions}
The research was performed by J.L. and I.K. The manuscript was written by J.L. and I.K. The numerical simulations were conducted by J.L.

\paragraph{Competing Interests}
The Authors declare no Competing Financial or Non-Financial Interests.

\bibliographystyle{IEEEtran} 
\bibliography{QuantumSpectralClustering_PRA.bib}

\end{document}